\documentclass[prd,nofootinbib,onecolumn]{revtex4}
\usepackage{amsmath}
\usepackage{amssymb}
\usepackage{graphics}
\usepackage{epsfig}

\newcommand\ba{\begin{eqnarray}}
\newcommand\ea{\end{eqnarray}}
\newcommand\nn{\nonumber}

\newcommand{\br}[1]{\left( #1 \right)}
\newcommand{\brs}[1]{\left[ #1 \right]}

\newcommand{\brm}[1]{\left| #1 \right|}
\newcommand{\GeV}{~\mbox{GeV}}
\newcommand{\MeV}{~\mbox{MeV}}
\newcommand{\KeV}{~\mbox{KeV}}

\begin{document}


\title{Radiative decays of pseudoscalar ($P$) and vector ($V$) mesons and
the process $e^+e^- \to \eta' \rho$}

\author{Yu.~M.~Bystritskiy}
\email{bystr@theor.jinr.ru}
\affiliation{Joint Institute for Nuclear Research, Dubna, Russia}

\author{E.~A.~Kuraev}
\email{kuraev@theor.jinr.ru}
\affiliation{Joint Institute for Nuclear Research, Dubna, Russia}

\author{M.~Se\v{c}ansk\'y}
\email{fyzimsec@savba.sk}
\affiliation{Institute of Physics, Slovak Academy of Sciences, Bratislava}

\author{M.~K.~Volkov}
\email{volkov@theor.jinr.ru}
\affiliation{Joint Institute for Nuclear Research, Dubna, Russia}

\begin{abstract}
Radiative decays of pseudoscalar and vector mesons are
calculated in the framework of the chiral Nambu-Jona-Lasinio (NJL) model.
We use the amplitude for triangle quark loops of anomalous type.
In evaluating these loop integrals we use two methods.
In the first one, we neglect the dependence of external momenta
by reproducing the Wess-Zumino-Witten terms
of effective chiral meson Lagrangian.
In the second method, we take into account the momentum
dependence of loop integrals
omitting their imaginary part. This makes it possible to allow
for quark confinement. As applied both the methods is in qualitative
agreement with each other and with experimental data.
The second method allows us to describe the electron-positron annihilation
with production of $\eta'$ and $\rho$ mesons in the center of mass energy
range from 1.6 to 3.5$\GeV$. The comparison with the recent experimental data
is presented.
\end{abstract}

\maketitle

\section{Introduction}

In the local NJL model \cite{Nambu:1961tp}, \cite{Volkov:1986zb}, \cite{Volkov:2006vq}
meson interactions are described in terms
of quark loops. If one neglects the
dependence of external momenta in the corresponding integrals,
the result will not violate chiral symmetry.
In this way we can reproduce effective
chiral Lagrangian corresponding to $U(3)\times U(3)$ symmetry
\cite{Gasiorowicz:1969kn}, \cite{Volkov:1986zb}, \cite{Ebert:1985kz}.
In this lagrangian strong interaction vertices are expressed
in terms of logarithmically divergent parts of corresponding loop
amplitudes. Radiative interactions of mesons are described in
terms of quark Feynman diagrams of anomalous type which do
not contain ultraviolet divergencies.
We use these anomalous quark loops for description of radiative
mesons decays.

However for production of pseudoscalar and vector mesons in
electron-positron collisions we should keep the
external momentum dependence of the amplitudes. In this case, we encounter
a serious problem of providing quark confinement condition.
To solve this problem, one usually uses a nonlocal version of the NJL model,
which involves the relevant formfactors for description of
the interaction between mesons and quarks. The choice of
formfactors leads to the functional unambiguity.
Among the different ways to introduce these formfactors we should mention
the QCD approach \cite{Arbuzov:2006ia} and the formfactors which
arises from the instanton model \cite{Plant:1997jr}. It is necessary to mention
the models suggested by G. Efimov \cite{Efimov:1993zg},
Yu. Simonov \cite{DiGiacomo:2000va}, Roberts and so on.

It should be noted that a different form of formfactors leads to
a different behavior of amplitudes in the physical region.
This becomes essential at large values of external momenta.
It is the reason why we apply rather a simple and rough
method which consists in exact calculation of amplitudes
and neglection of their imaginary parts to avoid the production
of free quarks. As a result, we obtain a rather satisfactory description
of radiative decay widths of vector and pseudoscalar mesons.
This fact allows us to hope that this approach can be applied to
describe the processes $e \bar e \to \eta' \rho$,
$e\bar e \to \eta' \pi^+\pi^-$ which
can be measured in a series of existing and planned experiments
with colliding electron-positron beams
\cite{Akhmetshin:2006sc}, \cite{Aubert:2007ef}, \cite{Aloisio:2002vm},
\cite{Ablikim:2006eg}.

For the problems of the last type we hope to obtain only qualitative
results in the center of mass energy range $1-3\GeV$.
In our approach we do not introduce any quark-meson formfactors.

Describing the decays of light mesons we ignore
the dependence of corresponding loop integrals of external momenta
-- Approximation I.
The problem of confinement in that approximation is automatically solved.
In the case of heavy mesons we keep exact external momentum dependence of
relevant loop amplitudes and neglect a possible imaginary part -- Approximation II.
Approximation II is used further for description of processes at electron-positron
colliders.

\section{Radiative decays of vector and pseudoscalar mesons}

For the description of interaction of mesons with quarks we
use the NJL model lagrangian \cite{Volkov:1986zb,Volkov:2006vq}:
\ba
    {\cal L}_{int} =
    \bar q \brs{ e Q \hat A +
    (i \gamma_5) \br{g_u \lambda_u \eta_u + g_s \lambda_s \eta_s} +
    \frac{g_\rho}{2} \br{ \lambda_3 \hat \rho_0 + \lambda_u \hat \omega + \lambda_s \hat \phi }
    } q,
    \label{QuarkMesonLagrangian}
\ea
where $\bar q= \br{\bar u,\bar d,\bar s}$ where $u$, $d$, $s$ are
the quark fields,
$Q=\mbox{diag}\br{2/3,-1/3,-1/3}$ is the quark charge matrix,
$\lambda_u=\br{\sqrt{2} \lambda_0+\lambda_8}/\sqrt{3}$,
$\lambda_s=\br{-\lambda_0+\sqrt{2}\lambda_8}/\sqrt{3}$
where $\lambda_i$ are the Gell-Mann matrices and
$\lambda_0 = \sqrt{2/3}~\mbox{diag}\br{1,1,1}$.
$g_u = m_u/f_\pi$, $g_s = m_s/f_s$ are the meson-quark coupling
constants which are evaluated by Goldberger-Treiman relation
($m_u = 263\MeV$, $m_s = 407\MeV$ are quark masses \cite{Radzhabov:2007wk},
and $f_\pi=92.4\MeV$ is the pion decay constant and $f_s = 1.3 f_\pi$).
$g_\rho=5.94$ is the $\rho \to 2\pi$ coupling constant.

Physical states of $\eta$ and $\eta'$ mesons are obtained after
taking into account of singlet-octet mixing
of $\eta_u$ and $\eta_s$ with
the angle $\theta=51.3^o$ \cite{Volkov:1998ax,Volkov:1986zb}:
\ba
    \eta &=& -\eta_u \sin\theta + \eta_s \cos\theta, \label{EtaEtaPrimeMixing}\\
    \eta' &=& \eta_u \cos\theta + \eta_s \sin\theta. \nn\\
\ea

We will consider the following processes below:
$\rho (\omega) \to \eta \gamma$, $\eta' \to \rho (\omega) \gamma$,
$\phi \to \eta (\eta') \gamma$.

The vector meson decay
\ba
V(p_1) \to \gamma(p_2) + P(p_3), \nn
\ea
is described by the amplitude of one loop with quark
(see Fig.~\ref{FigQuarkTriangle}):
\begin{figure}
\includegraphics{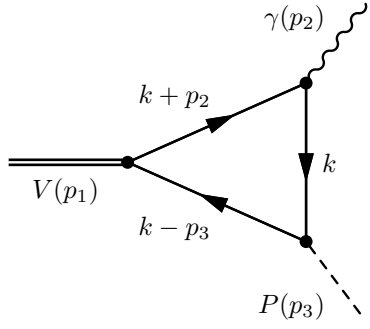}
\caption{The radiative decay amplitude (\ref{QuarkTriangleAmplitude}).
\label{FigQuarkTriangle}}
\end{figure}
\ba
    M_{V\to P\gamma} = \frac{i}{\br{2\pi}^2}~e~g_P~g_V~C_{PV}~
    M_q~J(p_1^2, 0, p_3^2) (e_1 e_2 p_1 p_2),
    \label{QuarkTriangleAmplitude}
\ea
where $(abcd) \equiv \varepsilon_{\alpha\beta\gamma\delta} a^\alpha b^\beta c^\gamma d^\delta$,
$g_V=g_\rho/2$, $g_P=g_u$ if light quarks go through the loop
and $g_P=g_s$ if strange quarks are involved;
$C_{PV}$ is the flavour-color multiplier corresponding to quark-meson
interaction, $C_{\eta\omega}=2\sin\theta$, $C_{\eta\rho}=6\sin\theta$,
$C_{\eta'\omega}=2\cos\theta$, $C_{\eta'\rho}=6\cos\theta$,
$C_{\eta\phi}=4\cos\theta$, $C_{\eta'\phi}=4\sin\theta$;
$M_q$ is the loop quark mass and
\ba
    J(p_1^2, p_2^2, p_3^2) &=&
    \mbox{Re}\br{
        \int \frac{dk}{i\pi^2}\frac{1}
        {\br{M_q^2 - k^2- i 0}\br{M_q^2 - (k+p_2)^2- i 0}\br{M_q^2 - (k-p_3)^2- i 0}}
    }
    =\nn\\
    &=&
    \mbox{Re}\br{
        \int_0^1 dx \int_0^{1-x} dy \frac{1}
        {M_q^2 - x y p_1^2 - y z p_2^2 - x z p_3^2 - i 0}
    },
    \label{ExactJ}
\ea
where $z = 1-x-y$.
In the heavy quark approximation (Approximation I) we obtain
\ba
J(p_1^2, p_2^2, p_3^2) = \frac{1}{2 M_q^2}. \label{ApproxiumateJ}
\ea
In this approximation a wide
set of decays of light mesons was described
in \cite{Volkov:1986zb} and the results were found to be
in good agreement with the experimental data.
The matrix element square can be written in the form:
\ba
    \brm{M_{V\to P\gamma}}^2 =
    \frac{e^2~g_P^2~g_V^2~C_{PV}^2}{\br{2\pi}^4}~
    \br{M_q~J(M_V^2, 0, M_P^2)}^2
    \brs{\frac{1}{2}\br{M_V^2-M_P^2}^2}.
\ea
The phase volume of the final state is:
\ba
    d\Phi_{P\gamma} =
    \frac{d^3 p_2}{2 E_2} \frac{d^3 p_3}{2 E_3}
    =
    \frac{1}{8\pi} \frac{M_V^2 - M_P^2}{M_V^2}.
\ea
And then the decay width reads as:
\ba
    \Gamma_{V\to P\gamma} =
    \frac{1}{3}
    \frac{\alpha}{2^7 \pi^4} \left(\frac{M_V^2-M_P^2}{M_V}\right)^3
    \left[g_P~g_V~C_{PV}~M_q~J(M_V^2, 0, M_P^2)\right]^2.
\ea
%
The relevant expression for radiative pseudoscalar meson decays $P \to V \gamma$
has the form:
\ba
    \Gamma_{P \to V\gamma} =
    \frac{\alpha}{2^7 \pi^4} \left(\frac{M_P^2-M_V^2}{M_P}\right)^3
    \left[g_P~g_V~C_{PV}~M_q~J(M_V^2, 0, M_P^2)\right]^2.
\ea
\begin{table}
\begin{tabular}{|l|r|r|r|}
\hline
Decay & Experiment & Approximation I (\ref{ApproxiumateJ}) &
Approximation II (\ref{ExactJ}) \\
\hline
$\rho \to \eta \gamma$ & $39.47$ & $65$ & $33.72$ \\
\hline
$\omega \to \eta \gamma$ & $4.07$ & $7.83$ & $4.16$ \\
\hline
$\eta' \to \rho \gamma$ & $59.68$ & $76.18$ & $41.09$ \\
\hline
$\eta' \to \omega \gamma$ & $6.15$ & $7.59$ & $4.04$ \\
\hline
$\phi \to \eta \gamma$ & $55.59$ & $71.01$ & $117.9$ \\
\hline
$\phi \to \eta' \gamma$ & $0.265$ & $0.497$ & $0.294$ \\
\hline
\end{tabular}
\caption{The table of radiative decays. The values of the widths are in $\KeV$.
Approximation I -- neglect of the external momentum dependence.
Approximation II -- the real part of exact loop integrals is taken into account.
\label{TableDecaysSummary}}
\end{table}
In Table \ref{TableDecaysSummary} we present the
theoretical results for both the methods -- Approximation I (\ref{ApproxiumateJ})
and Approximation II (\ref{ExactJ}) -- and compare them with the
relevant experimental data.

In particular, we would like to note that the ratio
$R_{th.} = \Gamma(\phi \to \eta' \gamma)/\Gamma(\phi \to \eta \gamma)=2.49\times10^{-3}$
is in qualitative agreement with the result of the KLOE collaboration
$R_{exp.}=\br{4.70\pm 0.47(\mbox{stat.})\pm 0.31(\mbox{syst.})}\times10^{-3}$
\cite{Aloisio:2002vm}.

\section{Associative production of pseudoscalar and vector mesons in
electron-positron annihilation}

The matrix elements of the processes of associative production of
pseudoscalar and vector mesons
\ba
e^+(p_+)+e^-(p_-) \to V(p_1) + P(p_3),
\ea
where $s=(p_++p_-)^2$, $p_\pm^2=m^2$, $p_1^2=M_V^2$, $p_3^2=M_P^2$,
in the lowest order of the QED coupling constant $\alpha$ have the form
(see Fig.~\ref{FigEEtoPV}):
\begin{figure}
\includegraphics{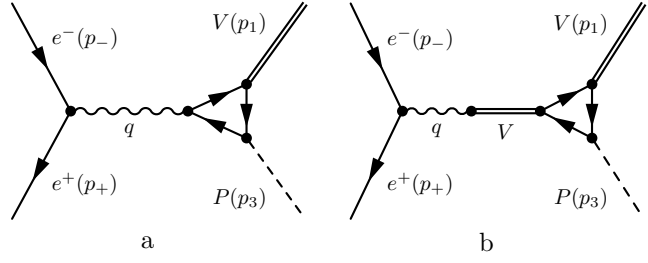}
\caption{The processes at electron-positron colliders.
\label{FigEEtoPV}}
\end{figure}
\ba
M^{PV}=i \frac{4\pi\alpha}{s}J^{em}_\mu J^{A\mu},
\ea
where the QED lepton current is $J^{em}_\mu=\bar{v}(p_+) \gamma_\mu u(p_-)$
and the anomalous current has the form
\ba
J^{A\mu}= \frac{g_P~g_V~C_{PV}}{\br{2\pi}^2}
\br{e_1 \mu p_1 p_2}~M_q~J(p_1^2, s, p_3^2),
\ea
where $p_++p_-= p_2 = p_1+p_3$ and $e_1$ is the polarization vector of the vector meson
(i.e. $(e_1 p_1)=0$).

The cross section built by general rules is
\ba
d\sigma=\frac{1}{8s}\sum |M^{PV}|^2 d\Phi_{PV},
\ea
where the phase volume of the final state has the form:
\ba
d\Phi_{PV}=\frac{d^3p_1}{2E_1}\frac{d^3p_3}{2E_3}\frac{1}{(2\pi)^2}\delta^4(p_++p_--p_1-p_3).
\ea
As we are concerned with the total cross section only we can use
the property of anomalous current gauge invariance and thus
rewrite the final state phase integral as
\ba
\sum \int J^A_\mu (J^A_\nu)^* d\Phi_{PV}=\frac{1}{3}\br{g_{\mu\nu}-\frac{q_\mu q_\nu}{q^2}}\int\brm{J^A_\mu}^2 d\Phi_{PV}.
\label{GaugeInvarianceTrick}
\ea
The second term in the braces does not give a contribution due to
gauge invariance of the lepton current $J^{em}_\mu$. The first term contribution is
proportional to
\ba
    \sum \br{e_1 \mu p_1 p_2}\br{e_1^* \mu p_1 p_2}
    =
    - \frac{1}{2} \lambda\br{s,M_P^2,M_V^2},
\ea
where $\lambda(a,b,c)=a^2+b^2+c^2-2ab-2ac-2bc$ is the well-known triangle function.
Thus, the quantity $\brm{J^A_\mu}^2$ in (\ref{GaugeInvarianceTrick})
does not depend on the vectors $p_1$ and $p_2$ themselves but only of
their squares $p_1^2=M_V^2$, $p_3^2=M_P^2$. This allows us to
calculate the integral over the final state phase volume
which, neglecting the masses of leptons, can be written as
\ba
    \int \frac{d^3p_1}{2E_1}\frac{d^3p_3}{2E_3} \delta^4(q-p_1-p_3)
    =
    \frac{\pi}{2}
    \frac{\lambda^{\frac{1}{2}}\br{s,M_P^2,M_V^2}}{s}.
\ea
Then the total cross section obtains the form:
\ba
    \sigma(s)=\frac{\alpha^2}{96\pi^3 s^3}
    \lambda^{\frac{3}{2}}\br{s,M_P^2,M_V^2}
    \brm{g_V~g_P~C_{PV}~M_q~J\br{M_P^2,M_V^2,s}}^2.
    \label{TotalCrossSection}
\ea
%
%
%
The differential cross section can be written as:
\ba
\frac{d\sigma}{d\Omega_2} = \sigma(s) \frac{3\br{1+\cos^2\theta}}{16\pi},
\ea
where $\theta$ is the center of mass angle between the
direction of 3-momenta of the initial electron $\vec p_-$ and the final vector
particle momentum direction $\vec p_1$.

\begin{figure}
\includegraphics[width=0.8\textwidth]{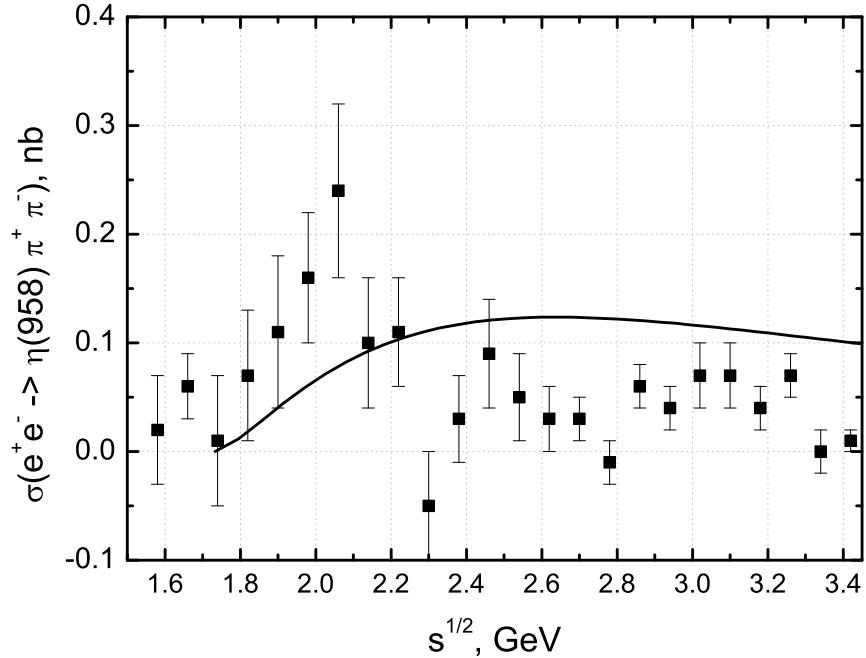}
\caption{The comparison of our result for the
$e^+e^- \to \eta' \pi^+ \pi^-$ with the BABAR-collaboration results for the
$e^+e^- \to \eta' \pi^+ \pi^-$ channel \cite{Aubert:2007ef}.
\label{FigBABARComparisonEtaPrimePiPi}}
\end{figure}

Let us consider the concrete process $e\bar e \to \eta' \rho$.
The expression for the total cross section (\ref{TotalCrossSection})
implied only contact Feynman diagram, i.e., Fig~\ref{FigEEtoPV} a.
Recalling the possible conversion of virtual
photon into vector mesons beyond resonances one must take into
account the diagrams presented on Fig~\ref{FigEEtoPV}, b. This leads to
the replacement of the factor $s^{-3}$ in (\ref{TotalCrossSection})
by the following one:
\ba
\frac{1}{s^3}
\to
\frac{1}{s^3}
\br{
1-\frac{1}{2\br{1-\frac{M_\rho^2}{s}}}
}^2.
\ea
The cross section of the process $e \bar e \to \eta'(950) \rho$ is drawn on
Fig.~\ref{FigBABARComparisonEtaPrimePiPi}, where the relevant experimental data
are also shown.
One can conclude that satisfactory agreement within the experimental errors is observed.

\begin{figure}
\includegraphics[width=0.8\textwidth]{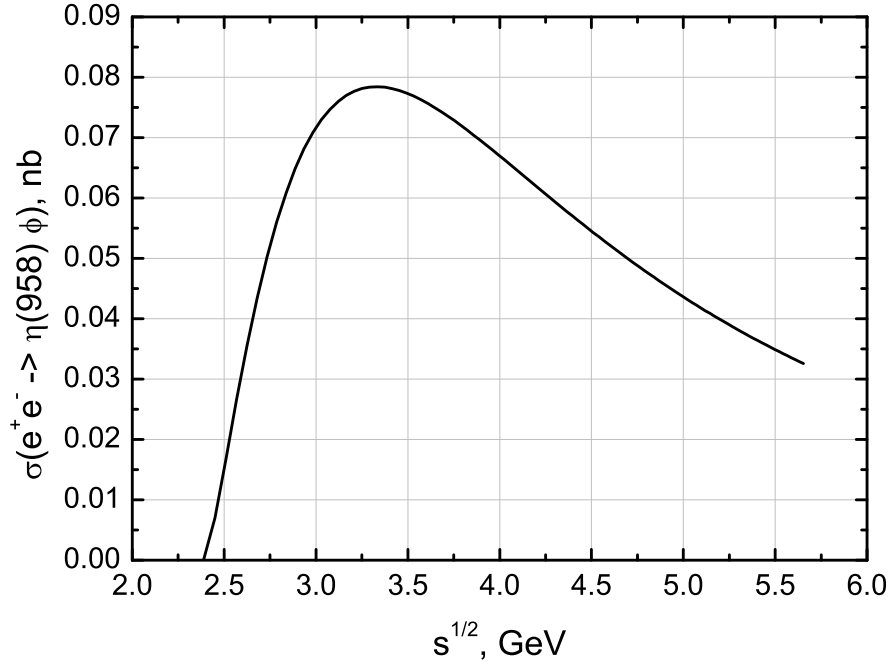}
\caption{The prediction for cross section $e^+e^- \to \eta' \phi$.
\label{FigEtaPrimePhi}}
\end{figure}

Let us now make a prediction for the process $e\bar e \to \eta' \phi$.
The relevant correction factor is
\ba
\frac{1}{s^3}
\to
\frac{1}{s^3}
\br{
1-\frac{1}{3\sqrt{2}\br{1-\frac{M_\phi^2}{s}}}
}^2.
\ea
Besides only $s$-quark loop works. The result is given in Fig.~\ref{FigEtaPrimePhi}.
%

\section{Conclusion}

In this paper, we investigated the radiative decays of vector and
pseudoscalar mesons described by the quark loops of anomalous type.

Let us note that in \cite{Bystritskiy:2007wq} we considered the process
$\phi\to f_0(980) \gamma$ within the same framework of the NJL model.
However, there the quark loop contribution was small enough and the main contribution
arose from terms of next order of $1/N_c$ expansion (where $N_c$ is the number of
colors) -- meson loops. In this paper, we have another situation:
meson loops are absent totaly and only quark loops of anomalous type give
a contribution to the amplitude of the process.

Both the approaches (Approximation I and Approximation II)
were considered. We show that the application
of the NJL model leads to rather satisfactory agreement with the modern
experimental data for the radiative decays. That allows us to use the Approximation II
to calculate the cross sections of associative vector and pseudoscalar
mesons production in the electron-positron annihilation channel in the lowest order
of electromagnetic constant. In the last case, the heavy virtual
photon converts to the set of pseudoscalar and vector mesons.
Two mechanisms must be taken into account: first one with the intermediate
virtual photon and the second one which contains the conversion of intermediate
photon into vector meson.
We give a comparison of our prediction
for the process $e\bar e \to \eta' \rho$ with the experimental data
of the BABAR collaboration.
For the process $e\bar e\to \eta' \phi$ the prediction is given for
future experimental data.

For comparison with the experiment the precision of our results
is worth mentioning.
We should like to notice that for the NJL model results precision is of an
order of 20-30~\%.

One of our important theoretical assumptions
is the absence of the imaginary part of relevant amplitudes.
The mechanism of elimination of the imaginary part is tightly
connected with the confinement nature and is not considered
here. We carry out the elimination "by hand" ("naive confinement").
It is to be noted, however, that if the
imaginary part is taken into account, the considerable disagreement
with the experimental data will occur in decay case.
For instance, if the imaginary part of the amplitude is taken into account,
the decay width of $\phi\to\eta'\gamma$ is
$\Gamma_{\phi\to\eta'\gamma}^{th.}=0.824\KeV$ while the experiment
gives $\Gamma_{\phi\to\eta'\gamma}^{exp.}=0.265\KeV$
(see Table~\ref{TableDecaysSummary}).

Concerning the singlet-octet angle mixing we use the additional
interaction of the t'Hooft type in lagrangian in the NJL model
\cite{Volkov:1998ax}, \cite{Volkov:1993jw}, \cite{Ebert:1994mf}.
This approach was widely used in literature
\cite{Klevansky:1992qe}, \cite{Vogl:1991qt}.

Let us note however that the alternative
solution of mixing angle problem was developed in
\cite{Ball:1995zv}, \cite{Escribano:2007cd}, \cite{Escribano:1999nh},
\cite{Escribano:2005qq}.
In this approach two mixing angles appear. The application of the both approach to
describe the decays of the pseudoscalar and vector mesons leads to the
similar results.

Our results for decays in
Approximation I are in agreement with the
ones obtained in \cite{Prades:1993ys}
(compare Table~\ref{TableDecaysSummary} and Table~2 in \cite{Prades:1993ys}).


%
%

\section{Acknowledgements}

The authors thank O.~V.~Teryaev, M.~A.~Ivanov, G.~V.~Efimov for criticism
and discussions, and N.~G.~Kornakov for his attention.
We acknowledge the support of
INTAS, grant (no. 05-1000008-8328).
The work was also supported in part by the Slovak Grant Agency for
Sciences VEGA, Grant No. 2/7116/27.


\end{document}